\documentclass[referee]{raa}           
\usepackage{graphicx,times}
\usepackage{natbib}
\usepackage{amssymb,amsmath}
\bibpunct{(}{)}{;}{a}{}{,}

\usepackage[a4paper=true,dvipdfm=true,pagebackref=true]{hyperref}
\hypersetup{pdftitle = The title of my PDF, pdfauthor = My name, pdfsubject= The subject, pdfkeywords = keyword1 keyword2 keyword3} 
\hypersetup{colorlinks = true, linkcolor = green, anchorcolor = red, citecolor = blue, filecolor = red, pagecolor = red, urlcolor = red}

\begin{document}

   \title{Revisiting the formation mechanism for coronal rain from previous studies}

 \volnopage{ {\bf 20XX} Vol.\ {\bf X} No. {\bf XX}, 000--000}
   \setcounter{page}{1}

   \author{Leping Li\inst{1,2}, Hardi Peter\inst{3}, Lakshmi Pradeep Chitta\inst{3}, Hongqiang Song\inst{4}}

   \institute{CAS Key Laboratory of Solar Activity, National Astronomical Observatories, Chinese Academy of Sciences, Beijing 100101, People's Republic of 
China; {\it lepingli@nao.cas.cn}\\
        \and
             University of Chinese Academy of Sciences, Beijing 100049, People's Republic of China \\
	\and
              Max Planck Institute for Solar System Research, D-37077 G\"{o}ttingen, Germany\\
\and 
              Shandong Provincial Key Laboratory of Optical Astronomy and Solar-Terrestrial Environment, and Institute of Space Sciences, Shandong University, Weihai, Shandong 264209, People's Republic of China\\
\vs \no
   {\small Received 20XX Month Day; accepted 20XX Month Day}
}

\abstract{Solar coronal rain is classified generally into two categories: flare-driven and quiescent coronal rain. The latter is observed to form along both closed and open magnetic field structures. 
%
%
Recently, we proposed that some of the quiescent coronal rain events, detected in the transition region and chromospheric diagnostics, along loop-like paths could be explained by the formation mechanism for quiescent coronal rain facilitated by interchange magnetic reconnection between open and closed field lines.
In this study, we revisited 38 coronal rain reports from the literature.
From these earlier works, we picked 15 quiescent coronal rain events out of the solar limb, mostly suggested to occur in active region closed loops due to thermal nonequilibrium, to scrutinize their formation mechanism.
Employing the extreme ultraviolet images and line-of-sight magnetograms, the evolution of the quiescent coronal rain events and their magnetic fields and context coronal structures is examined. 
We find that 6, comprising 40\%, of the 15 quiescent coronal rain events could be totally or partially interpreted by the formation mechanism for quiescent coronal rain along open structures facilitated by interchange reconnection.
The results suggest that the quiescent coronal rain facilitated by interchange reconnection between open and closed field lines deserves more attention.
}

   \authorrunning{L. P. Li et al. }            
   \titlerunning{Revisiting coronal rain}  
   \maketitle

%
\section{Introduction}           
\label{sec:int}

Solar coronal rain is a well-known and common observed phenomenon in the solar atmosphere \citep{2001SoPh..198..325S, 2012SoPh..280..457A, 2014ApJ...797...36S, 2019ApJ...884...34L, 2020ApJ...905...26L}. 
It is a most intriguing feature due to its link to the coronal heating and magnetic field, and has been widely investigated \cite[see a review in][and references therein]{2020PPCF...62a4016A}. 
Coronal rain was first discovered in the 1970s \citep{1970PASJ...22..405K, 1972SoPh...25..413L}, and defined observationally as cool ($\sim$10$^{4}$ K) and dense (10$^{10}$-10$^{11}$ cm$^{-3}$) blob-like material that forms in the hot (a few MK) corona in a timescale of minutes, and then falls to the solar surface along curved loop-like paths \citep{2012ApJ...745..152A}. 
It hence plays a fundamental role in the mass cycle between the hot, tenuous corona and cool, dense chromosphere \citep{2018ApJ...864L...4L, 2019ApJ...884...34L, 2020ApJ...905...26L, 2021ApJ...910..82L, 2019AA...630A.123K}. 

Coronal rain is mostly observed out of the solar limb as plasma emission in chromospheric lines, e.g., H$\alpha$ and Ca II H, and transition region lines, e.g., Si IV, or usually appears on the disk as absorption in chromospheric lines and extreme ultraviolet (EUV) wavelength channels \citep{2004AA...415.1141D, 2005AA...443..319D, 2010ApJ...716..154A, 2012SoPh..280..457A, 2015ApJ...806...81A}. 
It is a multithermal structure with temperatures ranging from the chromospheric (a few 10$^{3}$ K) up to transition region ($\sim$10$^{5}$ K) temperatures, that generally represents plasma cooling from even higher temperatures over million degrees Kelvin \citep{2020PPCF...62a4016A}. 
Coronal rain blobs have lengths in the range from a few hundred km to 1-20 Mm, with peak number values of 0.7-1\,Mm \citep{2012ApJ...745..152A}, and widths of 150-300 km \citep{2015ApJ...806...81A}. 
The material falls to the surface with speeds ranging from a few 10 to a few 100 km s$^{-1}$, with the peak at 80-100 km s$^{-1}$ \citep{2001SoPh..198..325S, 2009RAA.....9.1368Z, 2012ApJ...745..152A}. 
The downward acceleration is less than or around 100 m s$^{-2}$, almost a third of the gravitational value \citep{2001SoPh..198..325S, 2004AA...415.1141D}. 
For this observed lack in acceleration, one possible explanation is that the gas pressure restructures the downstream of coronal rain \citep{2014ApJ...784...21O, 2020PPCF...62a4016A}.   

Depending on its relation with the flare, coronal rain is classified generally into two categories: flare-driven and quiescent coronal rain \citep{2020PPCF...62a4016A}. 
In addition, according to the magnetic field structures in which it forms \citep{2020ApJ...905...26L, 2021ApJ...910..82L}, quiescent coronal rain is sorted into two kinds: one takes place along the non-flaring active region (AR) closed coronal loops \citep{2001SoPh..198..325S, 2010ApJ...716..154A}, and the other occurs along the open magnetic structures in both quiet Sun regions \citep{2018ApJ...864L...4L} and ARs \citep{2019ApJ...874L..33M}. 
For these three kinds of coronal rain, different formation mechanisms (FMs) have been widely investigated \citep[e.g.,][]{2003A&A...411..605M, 2004A&A...424..289M, 2014ApJ...797...36S, 2016ApJ...833..184S, 2018ApJ...864L...4L, 2019ApJ...884...34L, 2020ApJ...905...26L, 2021ApJ...910..82L, 2019ApJ...874L..33M, 2020PPCF...62a4016A}.

The flare-driven coronal rain frequently appears in post-flare loops \citep{2003ApJ...586.1417B, 2014ApJ...797...36S, 2016ApJ...833..184S, 2017ApJ...842...15L}. 
For the FM of flare-driven coronal rain (labelled as FM-I in Table\,\ref{tab}), the thermal conduction front and beamed nonthermal particles, produced during the flare, heat the chromospheric material quickly, leading to the chromospheric evaporation and then the greater mass loading of the post-flare loops \citep{2009ApJ...690..347L, 2016ApJ...827...27Z, 2017ApJ...841L...9L, 2018ApJ...856...34T}. 
At the top of post-flare loops, when radiative losses exceed heating input, thermal instability occurs \citep{1953ApJ...117..431P, 1965ApJ...142..531F}. 
Due to thermal instability, the hot evaporated plasma cools and condenses quickly. 
The cool plasma condensation then falls to the surface, under the effect of gravity, along the post-flare loops as flare-driven coronal rain \citep{2014ApJ...797...36S, 2016ApJ...833..184S}.

For the FM of quiescent coronal rain along AR closed loops (labelled as FM-II-1 in Table\,\ref{tab}) \citep{2003A&A...411..605M, 2004A&A...424..289M}, the heating events, concentrated at (near) the loop footpoints, result in the chromospheric evaporation and direct mass ejections into the loops \citep{2006ApJ...647.1452P, 2015AA...583A.109L}. 
The loops hence become hotter and denser quickly, and then cool and condense rapidly due to thermal nonequilibrium \citep{2010ApJ...716..154A}. 
Subsequently, under the effect of gravity, the cool condensation falls from the corona to the surface along one or both legs of the loops as quiescent coronal rain \citep{2020PPCF...62a4016A}. 
Moreover, along a coronal loop, formation of quiescent coronal rain that may be triggered by impulsive heating associated with magnetic reconnection was reported \citep{2019AA...630A.123K}. 
Same as the flare-driven coronal rain, the quiescent coronal rain here forms along closed magnetic field lines. 
However, different from the flare-driven coronal rain, the quiescent coronal rain is a recurring (quasi-periodic) phenomenon with an occurrence interval of several hours \citep{2012ApJ...745..152A, 2016ApJ...827...39K, 2018ApJ...853..176A, 2020AA...633A..11F}. 
This periodicity of quiescent coronal rain may be caused by thermal nonequilibrium \citep{2018ApJ...853..176A, 2020AA...633A..11F}.

Recently, a new FM for quiescent coronal rain along open field lines facilitated by interchange reconnection (labelled as FM-II-2 in Table\,\ref{tab}) was proposed \citep{2018ApJ...864L...4L}. 
In this FM, the curved higher-lying open structures move down toward the surface, and reconnect with the lower-lying closed loops, resulting in the formation of a magnetic dip in the former. 
The newly reconnected closed loops and open structures then appear, and retract away from the reconnection region. 
The coronal plasma surrounding the dip of higher-lying open structures converges into the dip, leading to the enhancement of plasma density in the dip. 
Triggered by the density enhancement, thermal instability occurs, and cooling and condensation of hot coronal plasma take place in the dip. 
A prominence thus forms \citep{1991ApJ...378..372A, 2015ApJS..219...17Y, 2020RAA....20..166C}, and then facilitates a speedup of the reconnection. 
Due to the successive reconnection, the condensation falls to the surface along the legs of newly reconnected closed loops and higher-lying open structures as quiescent coronal rain \citep{2018ApJ...864L...4L}. 
 
No associated flare and nonthermal emission is detected during the long-term reconnection process. 
Most of the magnetic energy may be converted into wave energy by reconnection as quasi-periodic fast propagating magnetoacoustic waves \citep{2018ApJ...868L..33L}. 
The repeated condensation, and subsequent coronal rain events along open structures facilitated by interchange reconnection were recently reported \citep{2019ApJ...884...34L}. Such events are observed with a mean occurrence interval of $\sim$6.6 hr \citep{2020ApJ...905...26L}. 
Different from the periodicity of quiescent coronal rain along AR closed loops, this repetition of quiescent coronal rain along open structures may be caused by interchange reconnection between open and closed structures \citep{2019ApJ...874L..33M, 2020ApJ...905...26L, 2021ApJ...910..82L}. 
Employing the data of Solar Dynamic Observatory \citep[SDO;][]{2012SoPh..275....3P} and Solar TErrestrial RElations Observatory \citep[STEREO;][]{2008SSRv..136....5K} A and B from different observing angles, the quiescent coronal rain events facilitated by interchange reconnection along open structures both on the disk and above the limb were analyzed \citep{2021ApJ...910..82L}.

The quiescent coronal rain events along open structures facilitated by interchange reconnection take place in quiet Sun regions \citep{2018ApJ...864L...4L, 2018ApJ...868L..33L, 2019ApJ...884...34L, 2020ApJ...905...26L, 2021ApJ...910..82L}. 
Employing the AR observations, ubiquitous quiescent coronal rain in the commonly occurring coronal magnetic topology of a large embedded bipole very near a coronal hole boundary, named as raining null point topologies (RNPT), was presented \citep{2019ApJ...874L..33M, 2021ApJ...907...41K}. 
The rain forms both within the legs of closed loops inside the separatrix dome and null and at the null and open outer spines, which may be caused separately by thermal nonequilibrium or/and via interchange reconnection \citep{2019ApJ...874L..33M}. 
The FM for quiescent coronal rain along open field lines facilitated by interchange reconnection between open and closed structures \citep{2018ApJ...864L...4L} is then further supported by \citet{2019ApJ...874L..33M}.

Using images and spectra from the Interface Region Imaging Spectrograph \citep[IRIS;][]{2014SoPh..289.2733D}, we analyzed a quiescent coronal rain event along curved loop-like paths using transition region and chromospheric diagnostics \citep{2020ApJ...905...26L}, which would usually fit the framework of heating-condensation cycles due to thermal nonequilibrium in closed loops. 
However, combing the observations of IRIS and Atmospheric Imaging Assembly \citep[AIA;][]{2012SoPh..275...17L} on board the SDO, we found that the IRIS coronal rain actually corresponds to the downflows of condensation facilitated by interchange reconnection between open and closed structures \citep{2020ApJ...905...26L}. 
This suggests that some of the quiescent coronal rain events with curved loop-like paths emitting in chromospheric and transition region lines could be explained by the FM for quiescent coronal rain along open structures facilitated by interchange reconnection. 

Observations of coronal rain, including the flare-driven and quiescent coronal rain, have been widely analyzed previously \citep[e.g.,][]{2001SoPh..198..325S, 2012SoPh..280..457A, 2014ApJ...797...36S, 2019ApJ...884...34L, 2020ApJ...905...26L}. 
Could some of these events, originally interpreted to occur along AR closed loops, be explained by the FM for quiescent coronal rain facilitated by interchange reconnection along open structures? 
To answer this question, we revisit these coronal rain events to check their FM by investigating the evolution of magnetic fields and context coronal structures. 
The method and observations are described in Section \ref{sec:obs}. 
The results and a summary and discussion are presented in Sections \ref{sec:res} and \ref{sec:sum}, respectively.

\section{Method and observations}
\label{sec:obs}

Making use of NASA's Astrophysics Data System (https://ui.adsabs.harvard.edu), we search for the refereed papers that analyzed the coronal rain events from the year of 2000 to 2021, with a key word  of coronal rain. 
Employing the JHelioviewer software \citep{2017A&A...606A..10M}, we revisit the evolution of the coronal rain events studied in these papers, and investigate the context coronal structures and magnetic fields and their evolution, to recheck the FM for coronal rain. 

In this work, we employ only the SDO/AIA 171, 131, and 304 \AA~images, with a time cadence of 12\,s and spatial sampling of 0.6\arcsec~pixel$^{-1}$, to study the evolution of the coronal rain events and their context coronal structures. 
Due to their lower time cadences and spatial samplings, EUV images from the Solar and Heliosphere Observatory (SOHO) and STEREO are not used.
Different AIA wavelength channels show plasma with different characteristic temperatures, e.g., 171 \AA~(Fe IX) peaks at $\sim$0.9 MK, 131 \AA~(Fe VIII and Fe XXI) peaks at $\sim$0.6 MK and $\sim$10 MK, and 304 \AA~(He II) peaks at $\sim$0.05 MK. 
Helioseismic and Magnetic Imager \citep[HMI;][]{2012SoPh..275..229S} line-of-sight (LOS) magnetograms on board the SDO, with a time cadence of 45\,s and spatial sampling of 0.5\arcsec~pixel$^{-1}$, and AIA 211 \AA~images are employed to analyze the evolution of  magnetic fields associated with the coronal rain events and their context coronal structures.
Here, in order to better show the evolution of the coronal rain events and their context coronal structures, some AIA 171, 304, and 211 \AA~images, e.g., in Figs. \ref{f:mr-101126}, \ref{f:mr-140501}-\ref{f:mf-140501}, \ref{f:mf-101031}-\ref{f:mf-120222}, and \ref{f:mf-151209}, have been enhanced using the Multiscale Gaussian Normalization (MGN) technique \citep{2014SoPh..289.2945M}.

\section{Results}
\label{sec:res}

For the coronal rain previously studied, 38 refereed papers are selected; see column 2 in Table\,\ref{tab} for details. 
Among them, the papers investigating the same coronal rain events are merged into the same rows; see No. 10, 11, 16, 27, and 30 in Table\,\ref{tab} for examples. 
In total, there are 32 rows of papers as listed in Table\,\ref{tab}. 
From these papers, general information of the analyzed coronal rain events is obtained. 
In Table\,\ref{tab}, column 3 shows the dates of coronal rain events, and columns 4 and 5 separately reveal the source regions of coronal rain and their heliographic coordinates. 
Column 6 denotes whether the coronal rain occurred on the disk or out of the limb. 
The on-disk coronal rain events are studied in 9 papers, including the flare-driven coronal rain (No. 13, 16, and 31), and the quiescent coronal rain along closed loops (No. 4, 16, 22, 24, and 26) and open structures (No. 1), respectively. 
Moreover, both the on-disk and off-limb coronal rain events were investigated in \citet{2011AA...532A..96K} and \citet{2021ApJ...910..82L}; see No. 1 and 16 in Table\,\ref{tab}. 
Column 7 shows the observations that were employed to study the coronal rain events; see respective papers for more details. 

Column 8 of Table\,\ref{tab} indicates the original FM for coronal rain proposed in the papers.
Here, the FM-I denotes the FM for flare-driven coronal rain, and the FM-II-1 and FM-II-2 mean the FM for quiescent coronal rain separately along the closed and open field lines; see Section\,\ref{sec:int} for more details. 
In \citet{2014ApJ...797...36S}, both the quiescent coronal rain along non-flaring AR closed loops and flare-driven coronal rain were analyzed; see No. 16 in Table\,\ref{tab}. 
Two kinds of quiescent coronal rain separately along the AR inner closed loops and outer open spines were reported in \citet{2019ApJ...874L..33M} and \citet{2021ApJ...907...41K}, see No. 2 and 8 in Table\,\ref{tab}. 
In the 38 papers, the flare-driven coronal rain events were investigated in 4 papers \citep{2003ApJ...586.1417B, 2014ApJ...797...36S, 2016ApJ...833..184S, 2017ApJ...842...15L}; see No. 13, 16, and 31 in Table\,\ref{tab}. 
The quiescent coronal rain events along open structures were studied in 7 papers \citep{2018ApJ...864L...4L, 2018ApJ...868L..33L, 2019ApJ...884...34L, 2020ApJ...905...26L, 2021ApJ...910..82L, 2019ApJ...874L..33M, 2021ApJ...907...41K}; see No. 1-3, 6, 8, and 10 in Table\,\ref{tab}, and those along closed loops were analyzed widely in 28 papers; see Table\,\ref{tab} for more details.
Furthermore, the FM for coronal rain was not clearly provided in \citet{2012ApJ...745L..21L} and \citet{2015ApJ...807....7R}; see No. 18 and 23 in Table\,\ref{tab}. 

The coronal rain events, that occurred before the launch of the SDO in 2010 February, have no associated AIA observation. 
They are therefore not revisited in this study. 
Column 9 of Table\,\ref{tab} shows whether the studied coronal rain events have associated SDO observations. 
Moreover, the context coronal structures of coronal rain are difficult to observe on the disk \citep{2021ApJ...910..82L}. 
The on-disk coronal rain events are thus not revisited either. 
Additionally, the quiescent coronal rain events that have been identified to occur along open structures as well as the flare-driven coronal rain events are also not revisited. 
At last, 15 coronal rain events, mostly suggested to form along non-flaring AR closed loops due to thermal nonequilibrium, are selected to recheck their FM; see No. 5, 7, 9, 11, 12, 14, 15, 17-21, 23, and 26 in Table\,\ref{tab}.
Among them two coronal rain events were studied in \citet{2015ApJ...806...81A}, see No. 19 in Table\,\ref{tab}.
After investigating the evolution of these 15 coronal rain events and their magnetic fields and context coronal structures, we find that 6 events could be totally or partially explained by the FM for quiescent coronal rain along open structures facilitated by interchange reconnection; see No. 11, 12, 18, 20, 23, and 26 in Table\,\ref{tab}.

\subsection{Coronal rain event on 2010 November 26}
\label{sec:101126}
In \citet{2012ApJ...745L..21L}, condensation of coronal plasma leading to the formation of a prominence at magnetic dips of a transequatorial loop system, connecting the positive trailing polarity of AR 11126 in the south and the diffuse negative polarity in the north, on 2010 November 26 was presented. 
In the south of the transequatorial loop, a coronal rain event at the dips of the AR loops was detected; see No. 23 in Table\,\ref{tab}. 
It slid downward along the underlying loops with curved paths \citep{2012ApJ...745L..21L}. 
The coronal rain event here took place in the non-flaring AR loops.
Its FM was, however, not clearly proposed. 

\begin{figure}
   \centering
  \includegraphics[width=\textwidth, angle=0]{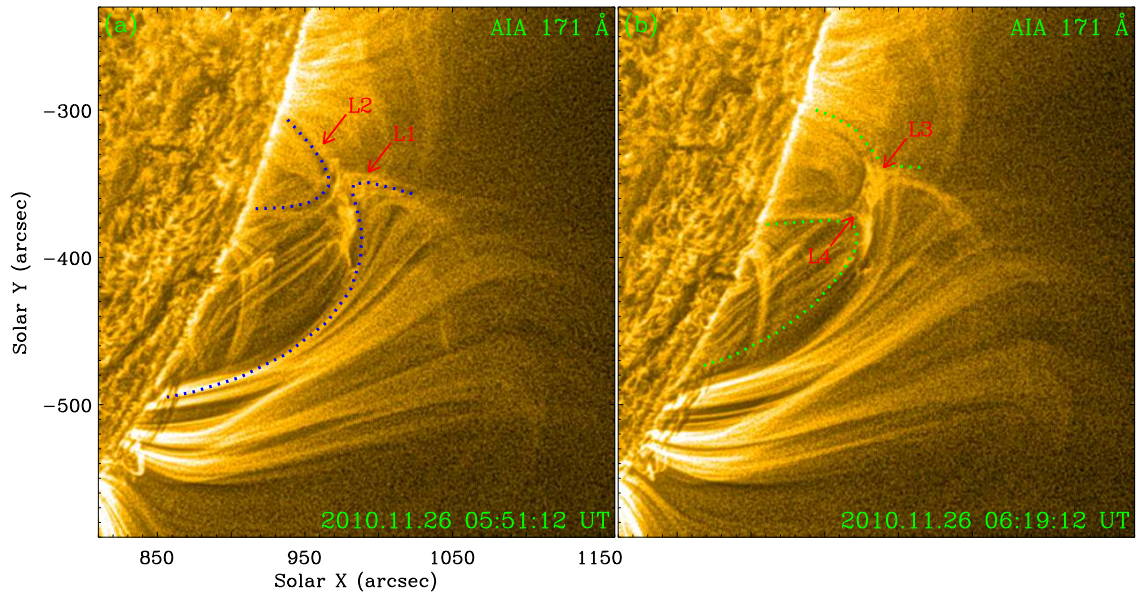}
   \caption{Magnetic reconnection between open and closed structures of the coronal rain event on 2010 November 26 \citep{2012ApJ...745L..21L}. 
(a) and (b) AIA 171 \AA~images that have been enhanced using the MGN technique. 
The blue dotted lines in (a) outline the higher-lying open structures L1 and lower-lying closed loops L2, respectively. 
The green dotted lines in (b) separately outline the newly reconnected open structures L3 and closed loops L4. 
An animation of the unannotated AIA 171\,\AA~images is available online. 
It covers 4 hr starting at 02:30 UT on 2010 November 26, with a time cadence of 1 minute. 
See Section \ref{sec:101126} for details. 
(An animation of this figure is available.)} 
   \label{f:mr-101126}
   \end{figure}

\begin{figure}
   \centering
  \includegraphics[width=\textwidth, angle=0]{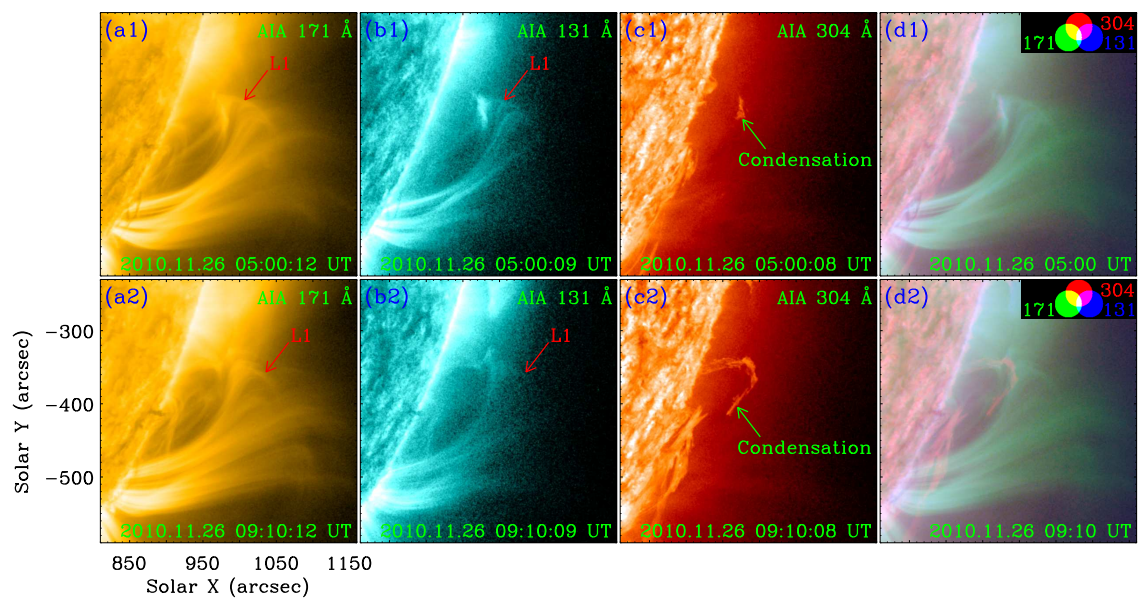}
   \caption{The coronal rain event on 2010 November 26 \citep{2012ApJ...745L..21L} observed by SDO. 
(a1) and (a2) SDO/AIA 171 \AA, (b1) and (b2) 131 \AA, and (c1) and (c2) 304 \AA~images, and (d1) and (d2) their composites.  
An animation of the unannotated AIA images is available online. It covers 8 hr starting at 02:30 UT on 2010 November 26, with a time cadence of 1 minute. 
See Section \ref{sec:101126} for details. 
(An animation of this figure is available.)} 
   \label{f:mr-cc-101126}
   \end{figure}

Using AIA 171, 131, and 304 \AA~images, we revisit the evolution of the coronal rain event in \citet{2012ApJ...745L..21L} and its context coronal structures. 
A set of higher-lying curved open structures rooting in AR 11126, labeled as L1 in Fig.\,\ref{f:mr-101126}(a), is observed in the AIA 171 \AA~channel.
The structures move down toward the surface, and reconnect with the lower-lying closed loops that are located in the quiet Sun region to the north of AR 11126, denoted by L2 in Fig.\,\ref{f:mr-101126}(a).  
The newly reconnected closed loops L4 and open structures L3 form in the process, outlined by the green dotted lines in Fig.\,\ref{f:mr-101126}(b); see the online animated version of Fig.\,\ref{f:mr-101126}.
During the reconnection process, a magnetic dip forms in the higher-lying open structures L1; see Fig.\,\ref{f:mr-cc-101126}(a1).
In the dip, bright emission sequentially appears in AIA 171 and 131 \AA~images; see Figs.\,\ref{f:mr-cc-101126}(a1) and (b1).
Here, the AIA 131 \AA~emission shows plasma with the lower characteristic temperature ($\sim$0.6 MK) of the AIA 131 \AA~channel, as no associated emission is detected in the AIA higher temperature channels, such as 94, 335, 211, and 193 \AA.
Cooling of coronal plasma hence happens in the dip of higher-lying open structures L1.

Thereafter, bright condensation forms in the dip of open structures L1 in AIA 304 \AA~images; see Fig.\,\ref{f:mr-cc-101126}(c1).
Figure\,\ref{f:mr-cc-101126}(d1) displays the composite of AIA 171, 131, and 304 \AA~images, illustrating the spatial relation among the AIA 171 \AA~structures, 131 \AA~emission, and 304 \AA~condensation.
Along with the successive reconnection, the condensation falls toward the surface along legs of the newly reconnected closed loops L4 and the higher-lying open structures L1 as coronal rain; see Figs.\,\ref{f:mr-cc-101126}(a2)-(d2) and the online animated version of Fig.\,\ref{f:mr-cc-101126}.
No associated flare is detected during the reconnection and condensation process.
The coronal rain thus belongs to quiescent coronal rain.
The evolution of the coronal rain event and its context coronal structures, i.e., the higher-lying open structures L1 and lower-lying closed loops L2, is identical to that in \citet{2018ApJ...864L...4L, 2018ApJ...868L..33L, 2019ApJ...884...34L, 2020ApJ...905...26L, 2021ApJ...910..82L}.
Therefore, we suggest that the coronal rain event in \citet{2012ApJ...745L..21L} could be explained by the FM for quiescent coronal rain facilitated by interchange reconnection between open and closed structures.

\subsection{Coronal rain event on 2014 May 1}
\label{sec:140501}
Employing IRIS and AIA observations, \citet{2015ApJ...807....7R} examined a small prominence eruption on 2014 May 1. 
They found evidence of reconnection between the erupting prominence magnetic field and the overlying field.
Prior to the prominence eruption, a coronal rain event was observed in the IRIS 1330 \AA~slit-jaw and AIA 304 \AA~images, forming a dome shape over the eruption site; see No. 18 in Table\,\ref{tab}.
The dome-shaped structure was proposed to be reminiscent of a fan-spine null point magnetic field configuration.
The FM for coronal rain was, however, not discussed.

We revisit the evolution of the coronal rain event in \citet{2015ApJ...807....7R} and its context coronal structures using AIA 171, 131, and 304 \AA~images.
In AIA 171 \AA~images, the higher-lying curved open structures, denoted by L1 in Fig.\,\ref{f:mr-140501}(a), are detected.
They move downward to the surface, and reconnect with the lower-lying closed loops L2, with the formation of a magnetic dip in the former; see the online animated version of Fig.\,\ref{f:mr-140501}.
The newly reconnected open structures L3 and closed loops L4 then form, outlined by the green dotted lines in Fig.\,\ref{f:mr-140501}(b).
Two sets of closed loops L2 and L4, two sets of open structures L1 and L3, and their evolution indicate a fan-spine null point magnetic field configuration of the AR; see Fig.\,\ref{f:mr-cc-140501}(a3).
Bright emission appears in AIA 171 and 131 \AA~images sequentially; see Figs.\,\ref{f:mr-cc-140501}(a1) and (b1), and coronal condensations then take place in AIA 304 \AA~images at the north edge of the dip of higher-lying open structures L1; enclosed by the blue ellipses in Figs.\,\ref{f:mr-cc-140501}(a1)-(c1).
This indicates the cooling and condensation process of coronal plasma in the dip.
The condensations fall to the surface northward along the leg of open structures L1 as coronal rain, and move southward into the dip along the structures L1 due to the motion of dip of structures L1 toward the surface; see Figs.\,\ref{f:mr-cc-140501}(a2)-(d2). 
Along with the successive reconnection, they also propagate along both legs of the newly reconnected closed loops L4 to the surface as coronal rain, outlined by the cyan dotted line in Fig.\,\ref{f:mr-cc-140501}(c3); see the online animated version of Fig.\,\ref{f:mr-cc-140501}.
Even though the coronal rain event occurs in the AR, all the results are almost the same as those in \citet{2018ApJ...864L...4L}, in which quiescent coronal rain is facilitated by interchange reconnection between open and closed structures. 

Using HMI LOS magnetograms, we investigate the long-term evolution of magnetic fields associated with the reconnection structures in several days before the coronal rain event on 2014 May 1, and notice that the reconnection structures root in AR 12044.
Figure\,\ref{f:mf-140501}(a) shows the magnetic fields of the AR on 2014 April 26, when they were located on the disk.
The negative magnetic fields, enclosed by the blue contours in Fig.\,\ref{f:mf-140501}(a), surrounded by the positive magnetic fields, outlined by the green contours in Fig.\,\ref{f:mf-140501}(a), are identified.
Here, most of the positive magnetic fields are located to the west and north of the negative ones, and several small positive magnetic field concentrations are located to the east and south. This supports the existence of  a fan-spine null point magnetic field configuration of the AR at coronal heights \citep{2019ApJ...871....4H, 2019ApJ...885L..11S, 2021ApJ...908..213L}, as shown in AIA images; see Fig.\,\ref{f:mr-cc-140501}(a3).
The open structures L1 root in the surrounding positive magnetic fields, located to the north of the central negative ones.
A simultaneous AIA 211 \AA~image in Fig.\,\ref{f:mf-140501}(b) illustrates the AR closed loops, connecting the central negative and surrounding positive magnetic fields marked respectively by the blue and green contours. 
It displays that the negative magnetic fields mostly connect with the positive magnetic fields located to the west and north, and rarely connect with those located to the south. 
The loops, connecting the negative magnetic fields and the positive ones that are located to the south, are denoted by the cyan solid arrows in Fig.\,\ref{f:mf-140501}(b).
To the north of the AR, a coronal hole with positive magnetic fields alone is detected; see Fig.\,\ref{f:mf-140501}(b).
The whole magnetic field configuration, including the embedded bipole and its nearby coronal hole, is consistent with the RNPT in \citet{2019ApJ...874L..33M}.
Moreover, evidence of interchange reconnection at the null point between the outer open spines and inner closed loops, suggested in \citet{2019ApJ...874L..33M}, is clearly found; see the online animated version of Fig.\,\ref{f:mr-140501}.
The coronal rain event in \citet{2015ApJ...807....7R} hence forms along outer open spines facilitated by interchange reconnection.
 
\subsection{Coronal rain event on 2010 October 31}
\label{sec:101031}
In \citet{2011AA...532A..96K}, the temporal evolution of loops in AR 11117 out of the northwest limb on 2010 October 31 was studied using AIA EUV images.
Coronal condensation took place occasionally in the AR loops, and flowed downward to the loop footpoints as coronal rain; see No. 26 in Table\,\ref{tab}.
It was interpreted in the framework of evaporation-condensation cycles due to thermal nonequilibrium, as quasi-periodic upflows from the loop footpoints in the AIA hot channel, e.g., 193 \AA, were detected just after the coronal rain \citep{2011AA...532A..96K}.

Employing AIA 171, 131, and 304 \AA~images, we revisit the evolution of the coronal rain event in \citet{2011AA...532A..96K} and its context coronal structures. 
A fan-spine null point magnetic field structure located in AR 11117 is observed; see Fig.\,\ref{f:mr-cc-101031}(a1). 
Bright emission appears sequentially in the AIA 171 and 131 \AA~channels; see Figs.\,\ref{f:mr-cc-101031}(a1) and (b1).
Thereafter, coronal condensation happens along the open outer spines, and flows downward to the surface along the fan surface as quiescent coronal rain in the AIA 304 \AA~channel; see Figs.\,\ref{f:mr-cc-101031}(c1) and (c2) and the online animated version of Fig.\,\ref{f:mr-cc-101031}.
Cooling and condensation of coronal plasma is evidently identified.

The long-term evolution of magnetic fields in AR 11117 in several days before the coronal rain event on 2010 October 31 is investigated using  HMI LOS magnetograms. 
On 2010 October 27, the AR magnetic fields on the disk are displayed in Fig.\,\ref{f:mf-101031}(a).
Central negative magnetic fields and their surrounding positive ones, marked by blue and green contours in Fig.\,\ref{f:mf-101031}(a), are identified. 
The embedded bipole constitutes a fan-spine null point magnetic field configuration, as displayed in AIA images; see Fig.\,\ref{f:mr-cc-101031}.
The open outer spines L1 in Fig.\,\ref{f:mr-cc-101031} root in the surrounding positive magnetic fields that are located to the east of the central negative ones.
A simultaneous AIA 211 \AA~image is illustrated in Fig.\,\ref{f:mf-101031}(b).
We overlay the central negative and surrounding positive magnetic fields in Fig.\,\ref{f:mf-101031}(a) on Fig.\,\ref{f:mf-101031}(b) as blue and green contours. 
Under the fan surface, inner closed loops connecting the central negative and surrounding positive magnetic fields, denoted separately by blue and green contours  in Fig.\,\ref{f:mf-101031}(b), are clearly detected.
More (less) loops connect the negative magnetic fields and the positive ones located to the east (west).
Moreover, a dark region, located to the east of AR, is identified, showing a coronal hole with only positive magnetic fields; see Fig.\,\ref{f:mf-101031}.
Similar to Section\,\ref{sec:140501}, an embedded bipole and its nearby coronal hole constitute the RNPT that was proposed in \citet{2019ApJ...874L..33M}.
The coronal rain event in \citet{2011AA...532A..96K} is thus facilitated by interchange reconnection between the outer open spines and inner closed loops at the null point.
 
\subsection{Coronal rain event on 2012 February 22}
\label{sec:120222}
Using AIA 171 and 304 \AA~images, \citet{2015AA...577A.136V} studied the temporal evolution of a coronal loop in which coronal rain took place in AR 11420 out of the east limb on 2012 February 22; see No. 20 in Table\,\ref{tab}. 
They found that the loop disappeared in the AIA 171 \AA~channel and appeared in the AIA 304 \AA~channel more than one hour later.
The rapid cooling of the loop was therefore observed.
The catastrophic cooling due to thermal nonequilibrium was accompanied by the formation of coronal rain in the shape of falling cold blobs along the loop, and was suggested to be responsible for the coronal rain formation. 
However, some blobs of coronal rain occurred below the cooling loop.
They could be a result of independent heating-condensation cycle occurring in the lower-lying loop that was unrelated to the higher-lying cooling loop, or may fall along the newly reconnected loop that was formed through reconnection \citep{2015AA...577A.136V}. 

The evolution of the coronal rain event in \citet{2015AA...577A.136V} and its context coronal structures is revisited by using AIA 171, 131, and 304 \AA~images. 
A set of curved open structures is observed in AIA 171 \AA~images, lasting for a long time; see L1 in Figs.\,\ref{f:mr-cc-120222}(a1) and (a2). 
They move downward to the surface, and reconnect with the lower-lying closed loops. 
The newly reconnected closed loops, outlined by a green dotted line in the green rectangle in Fig.\,\ref{f:mr-cc-120222}(a2), and open structures then form; see the online animated version of Fig.\,\ref{f:mr-cc-120222}.
Bright emission appears sequentially in AIA 171 and 131 \AA~images; see Figs.\,\ref{f:mr-cc-120222}(a1)-(a2) and (b1)-(b2).
Magnetic dips form in the open structures L1; see Figs.\,\ref{f:mr-cc-120222}(a1) and (a2), in which bright condensations occur in AIA 304 \AA~images; see Fig.\,\ref{f:mr-cc-120222}(c1).
The condensations fall toward the surface along the leg of open structures L1 and/or the north leg of the newly reconnected closed loops as quiescent coronal rain; see Fig.\,\ref{f:mr-cc-120222}(c2).

We investigate the evolution of associated magnetic fields in several days after the coronal rain event on 2012 February 22, employing the HMI LOS magnetograms.  
The magnetic fields on 2012 February 25, when they were located on the disk, are displayed in Fig.\,\ref{f:mf-120222}(a).
They belong actually to a plage region, a   few days later classified as AR 11426, rather than the AR 11420, that was located on the west limb.
In the plage region,  positive and  negative magnetic fields are observed,  enclosed separately by the green and blue contours in Fig.\,\ref{f:mf-120222}(a).
The open structures L1 root in the east of the negative magnetic fields that are located to the north of the  positive ones.
Figure\,\ref{f:mf-120222}(b) shows a simultaneous AIA 211 \AA~image, in which the green and blue contours  outline the  positive and  negative magnetic fields at the loop footpoints, respectively.
 Closed loops connecting the  positive and  negative magnetic fields are detected; see Fig.\,\ref{f:mf-120222}(b).
A coronal hole, with negative magnetic fields alone, is located to the west of the plage region; see Fig.\,\ref{f:mf-120222}.
The magnetic fields and coronal structures of the plage and  their nearby coronal hole  seem to be consistent with those in Fig. 2 of \citet{2019ApJ...874L..33M} that were suggested to show the fan-spine null point magnetic field topology. However, different from Sections\,\ref{sec:140501} and \ref{sec:101031}, the embedded bipole is hardly identified in  the plage region.
The coronal rain event in \citet{2015AA...577A.136V}  occurring along open structures may be  facilitated by interchange reconnection between open and closed structures in the quiet Sun region, as suggested in \citet{2018ApJ...864L...4L}.
Some coronal rain blobs occurring below the cooling loop then could be easily explained.

\subsection{Coronal rain event on 2012 July 24}
\label{sec:120724}
In \citet{2018ApJ...853..176A}, the long-term evolution of a system of large transequatorial loops and coronal rain in the loops above the east limb during 2012 July 23-25 was analyzed employing AIA EUV images; see No. 12 in Table\,\ref{tab}.
The periodic coronal rain was co-spatial and was in phase with the periodic intensity pulsations of transequatorial loops. 
They were suggested to be two manifestations of the same physical process: evaporation-condensation cycles resulting from a state of thermal nonequilibrium \citep{2018ApJ...853..176A}.

Using AIA 171, 131, and 304 \AA~images, we revisit the evolution of the coronal rain event in \citet{2018ApJ...853..176A} and its context coronal structures, and find a set of curved open structures located in the south of the transequatorial loops in the AIA 171 \AA~channel; see L1 in Fig.\,\ref{f:mr-cc-120724}(a1).  
The open structures L1 move down toward the surface, and form magnetic dips; see the online animated version of Fig.\,\ref{f:mr-cc-120724}. 
In the dips, bright emission appears in AIA 171 and 131 \AA~images sequentially; see Figs.\,\ref{f:mr-cc-120724}(a1) and (b1).
In the AIA 304 \AA~channel, condensations of coronal plasma subsequently happen in the dips, and fall to the surface along the leg of open structures L1 as quiescent coronal rain; see Fig.\,\ref{f:mr-cc-120724}(c1).
Cooling and condensations of coronal plasma are clearly identified along the open structures.

If we would consider the AIA 304 \AA~observations solely, the coronal rain seems to form totally in the transequatorial closed loops due to thermal nonequilibrium.
However, combining the evolution of coronal rain and its context coronal structures, we notice that part of the coronal rain appears along the open structures L1, that are superimposed on the transequatorial closed loops; see Fig.\,\ref{f:mr-cc-120724}.
Recently, employing observations of the SDO and STEREO from different viewing angles, \citet{2019ApJ...884...34L} suggested that the downward motion of higher-lying open structures, in which quiescent coronal rain occurs, represents the reconnection process between the higher-lying open structures and lower-lying closed loops, even though the whole reconnection magnetic structure is not totally detected.
We hence propose that part of the coronal rain event in \citet{2018ApJ...853..176A} corresponds to the downflows of condensations facilitated by interchange reconnection between open and closed structures.
Moreover, another similar coronal rain event along the open structures L1 took place late on 2012 July 24; see Figs.\,\ref{f:mr-cc-120724}(a2)-(d2) and the online animated version of Fig.\,\ref{f:mr-cc-120724}.
The coronal rain along open structures therefore occurs repeatedly, similar to those reported previously in \citet{2019ApJ...884...34L, 2020ApJ...905...26L, 2021ApJ...910..82L}.
 
\subsection{Coronal rain event on 2015 December 9}
\label{sec:151209}
\citet{2017SoPh..292..132S, 2018ApJ...865...31S} investigated the evolution of loops in AR 12468, that was located on the southeast limb, with transequatorial linkage to a distributed plage region in the northern hemisphere on 2015 December 9. 
Coronal rain originating from the transequatorial loops and from the magnetic structures above the AR was evidently detected; see No. 11 in Table\,\ref{tab}. 
Similar results to \citet{2018ApJ...853..176A} are obtained; see Section\,\ref{sec:120724}, in \citet{2017SoPh..292..132S, 2018ApJ...865...31S}.

The long-term evolution of the coronal rain event in \citet{2017SoPh..292..132S, 2018ApJ...865...31S} and its context coronal structures is revisited, using AIA 171, 131, and 304 \AA~images.
In the south of the transequatorial loops, a set of curved open structures is identified in AIA 171 \AA~images, denoted by L1 in Fig.\,\ref{f:mr-cc-151209}(a1). 
They move down toward the surface.
In the AIA 171 and 131 \AA~channels, bright emission sequentially appears; see Figs.\,\ref{f:mr-cc-151209}(a1) and (b1).
In the AIA 304 \AA~channel, coronal condensations then take place, and fall along the leg of open structures L1 as quiescent coronal rain; see Figs.\,\ref{f:mr-cc-151209}(c1) and (c2).
These results indicate the cooling and condensations of coronal plasma along the open structures.
    
The long-term evolution of magnetic fields associated with the open structures L1 in several days after the coronal rain event on 2015 December 9 is analyzed.
Figure\,\ref{f:mf-151209}(a) displays the magnetic fields of the AR on 2015 December 14, when the AR was located on the disk.
Central negative magnetic fields surrounded by positive ones, enclosed separately by the blue and green contours, are observed, indicating a fan-spine null point topology.
The open structures L1 root in the positive magnetic fields, located to the south of the central negative ones.
A simultaneous AIA 211 \AA~image is illustrated in Fig.\,\ref{f:mf-151209}(b).
Closed loops, connecting the central negative and surrounding positive magnetic fields; see the blue and green contours  in Fig.\,\ref{f:mf-151209}(b), under the fan surface are identified.
A coronal hole, with only positive magnetic fields, is located to the southwest of the AR.
Similar to Sections\,\ref{sec:140501} and \ref{sec:101031}, the fan-spine null point topology and its nearby coronal hole constitute the RNPT in \citet{2019ApJ...874L..33M}.
The coronal rain along open structures L1 is hence facilitated by interchange reconnection at the null point between the outer open spines and inner closed loops.
It is superimposed on the coronal rain originating from the transequatorial closed loops.
Similar to Section\,\ref{sec:120724}, we propose that part of the coronal rain event in \citet{2017SoPh..292..132S, 2018ApJ...865...31S} could be explained by the coronal rain FM along open structures facilitated by interchange reconnection.

\section{Summary and discussion}
\label{sec:sum}
Employing NASA's Astrophysics Data System, 38 refereed papers, analyzing the coronal rain events previously from 2000 to 2021, are selected; see Section\,\ref{sec:res}.
Among them, 4 papers studied the flare-driven coronal rain, and 
28 and 7 papers investigated the quiescent coronal rain respectively along the closed and open field lines; see Table\,\ref{tab}.
After 2010, 15 off-limb quiescent coronal rain events, mostly suggested to take place due to thermal nonequilibrium along AR closed loops, are chosen to revisit the FM for coronal rain; see Section\,\ref{sec:res}.
Using AIA 171, 131, 304, and 211 \AA~images and HMI LOS magnetograms on board the SDO, we investigate the evolution of the coronal rain events and their magnetic fields and context coronal structures.
Among the 15 quiescent coronal rain events, 6 ones could be totally/partially explained by the FM for quiescent coronal rain along open structures facilitated by interchange reconnection.

Some of the quiescent coronal rain events, explained originally by thermal nonequilibrium in closed loops driven by footpoint-concentrated heating events, could be interpreted by thermal instability facilitated by interchange reconnection between open and closed structures.
Recently, employing IRIS and AIA observations, \citet{2020ApJ...905...26L} found that some of the coronal rain events in transition region and chromospheric lines along curved loop-like paths correspond to the downflows of coronal condensations along open structures facilitated by interchange reconnection.
In this study, 6 of 15, comprising 40\%, selected quiescent coronal rain events out of the limb, originally explained mostly in the framework of evaporation-condensation cycles due to thermal nonequilibrium along the AR closed loops, in fact, form totally and/or partially along the open structures; see Sections\,\ref{sec:101126}-\ref{sec:151209}.
As the 15 quiescent coronal rain events revisited here take place generally in the ARs, and lots of quiescent coronal rain events along open structures occur in the quiet Sun regions \citep{2018ApJ...864L...4L, 2018ApJ...868L..33L, 2019ApJ...884...34L, 2020ApJ...905...26L, 2021ApJ...910..82L}, one can imagine that more quiescent coronal rain events may form along open structures.
The quiescent coronal rain along open structures facilitated by interchange reconnection between open and closed structures thus deserves more attention.

\begin{figure}
   \centering
  \includegraphics[width=\textwidth, angle=0]{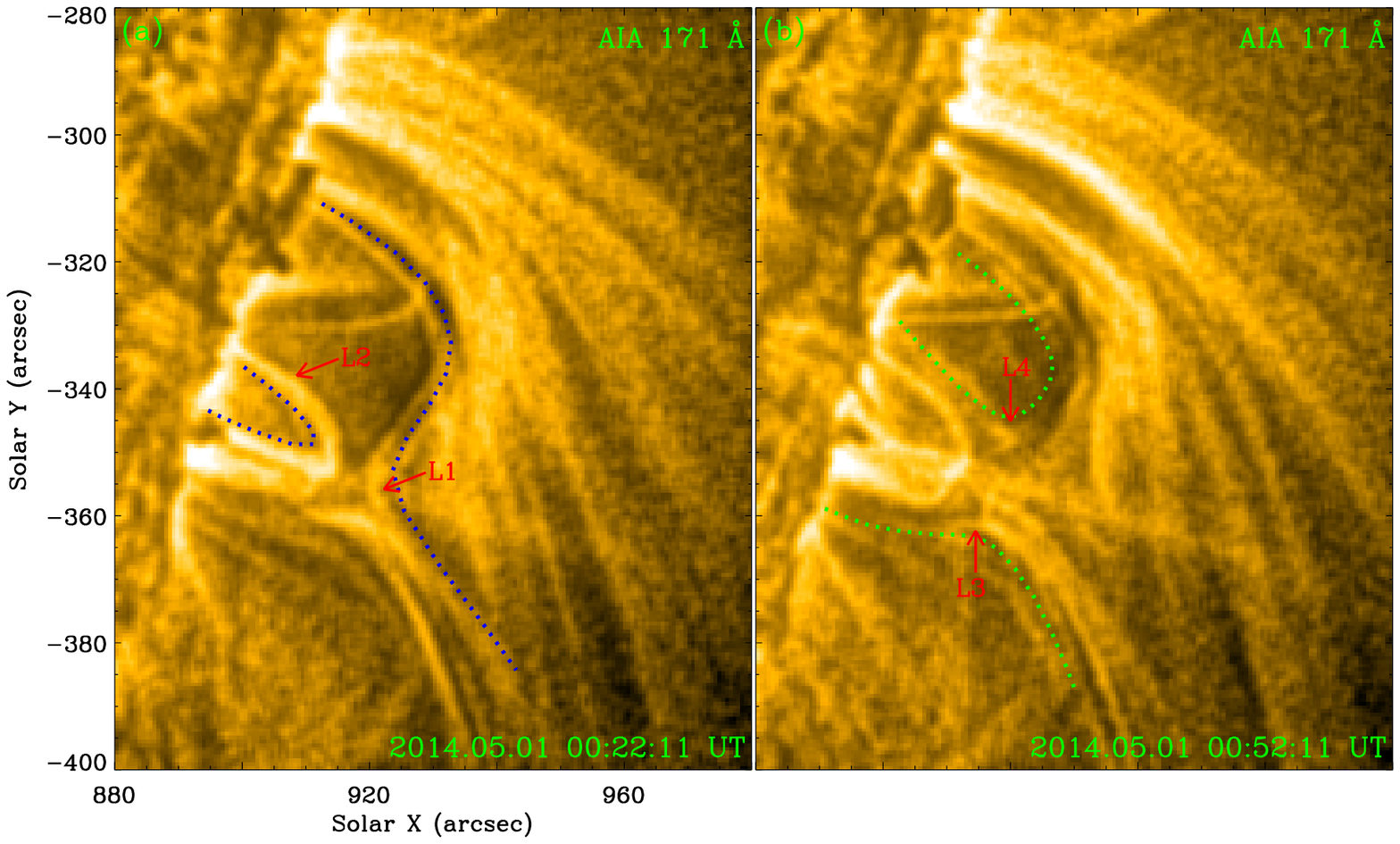}
   \caption{Similar to Fig.\,\ref{f:mr-101126}, but for the coronal rain event on 2014 May 1 \citep{2015ApJ...807....7R}.
An animation of the unannotated AIA 171\,\AA~images is available online. 
It covers 1 hr starting at 00:00 UT on 2014 May 1, with a time cadence of 1 minute. 
See Section \ref{sec:140501} for details. 
(An animation of this figure is available.)} 
   \label{f:mr-140501}
   \end{figure}

\begin{figure}
   \centering
  \includegraphics[width=\textwidth, angle=0]{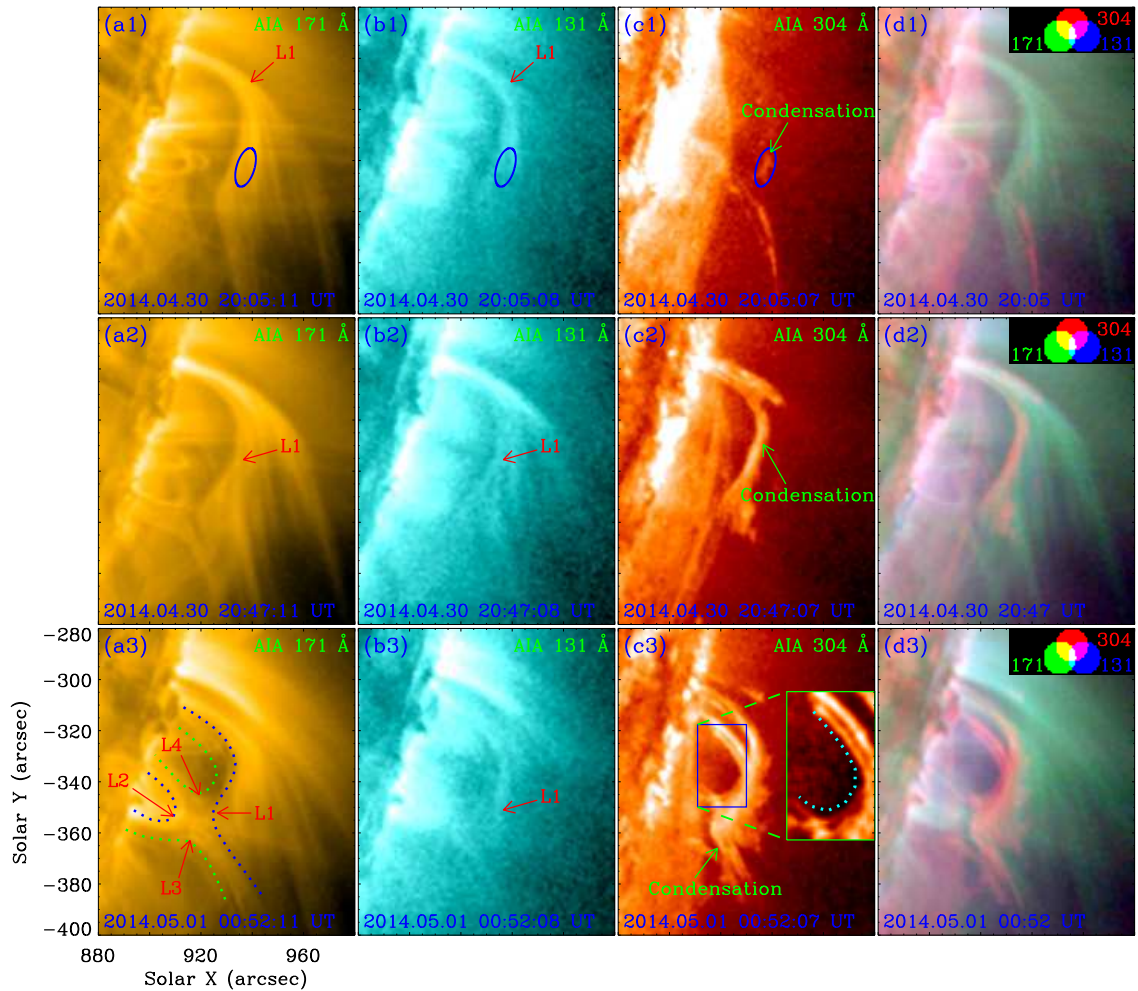}
   \caption{Similar to Fig. \ref{f:mr-cc-101126}, but for the coronal rain event on 2014 May 1 \citep{2015ApJ...807....7R}. 
The blue ellipses in (a1)-(c1) enclose the condensation of coronal plasma in (c1).
Same as in Fig.\,\ref{f:mr-140501}, the blue and  green dotted lines in (a3) separately outline the reconnecting and newly reconnected closed and open structures L2 and L1, and L4 and L3. 
In (c3), the AIA 304 \AA~image in the blue rectangle is enlarged in the green rectangle, and has been enhanced using the MGN technique.
The cyan dotted line in the green rectangle in (c3) outlines the coronal rain falling along the newly reconnected closed loops L4.
An animation of the unannotated AIA images is available online. 
It covers 6 hr starting at 20:00 UT on 2014 April 30, with a time cadence of 1 minute. 
See Section \ref{sec:140501} for details. 
(An animation of this figure is available.)} 
   \label{f:mr-cc-140501}
   \end{figure}

\begin{figure}
   \centering
  \includegraphics[width=\textwidth, angle=0]{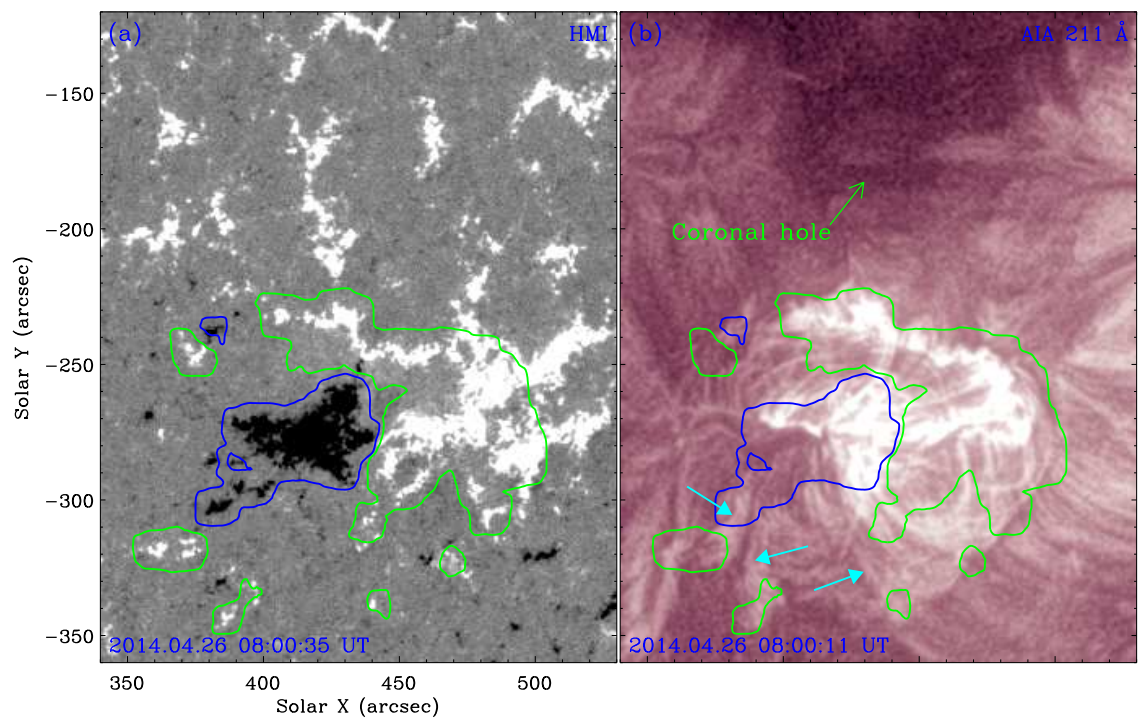}
   \caption{On-disk magnetic fields and context coronal structures observed by SDO on April 26 of the coronal rain event on May 1 \citep{2015ApJ...807....7R},  2014. 
(a) A SDO/HMI LOS magnetogram, and (b) an AIA 211 \AA~image. 
Here, the AIA 211\,\AA~image in (b) has been enhanced using the MGN technique.
The blue and green contours in (a) and (b)  enclose the  negative and  positive magnetic fields at the footpoints of AR loops, respectively. 
The cyan solid arrows in (b) denote the loops connecting the negative magnetic fields and the positive ones that are located to the south.
See Section \ref{sec:140501} for details.} 
   \label{f:mf-140501}
   \end{figure}

\begin{figure} 
   \centering
   \includegraphics[width=\textwidth, angle=0]{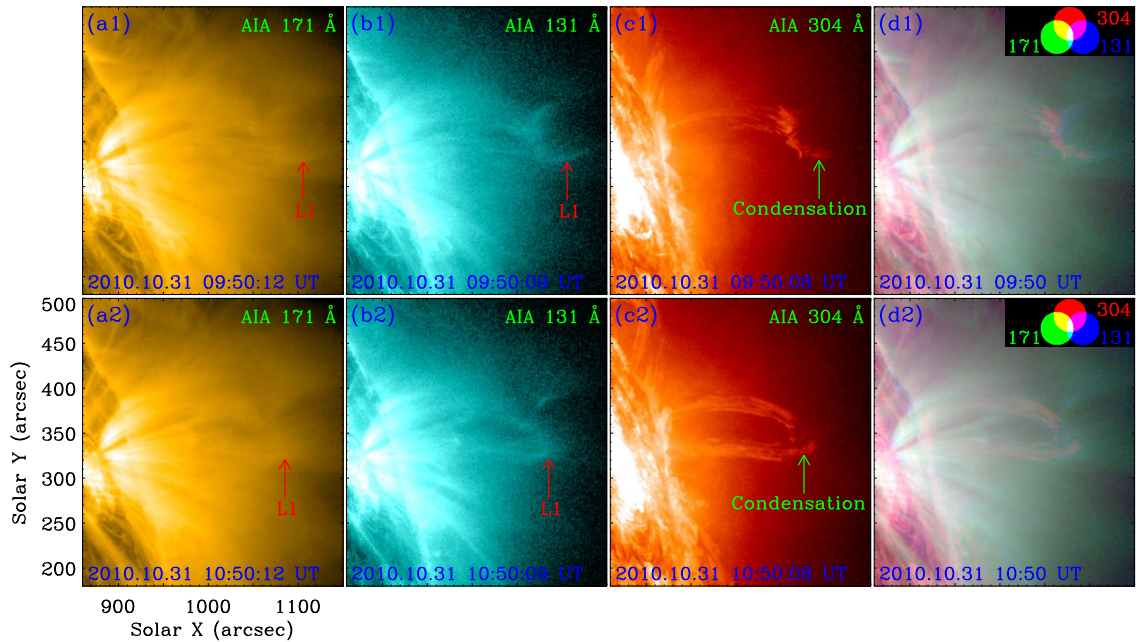}
   \caption{Similar to Fig. \ref{f:mr-cc-101126}, but for the coronal rain event on 2010 October 31 \citep{2011AA...532A..96K}. 
An animation of the unannotated AIA images is available online. 
It covers $\sim$3.3 hr starting at 09:45 UT on 2010 October 31, with a time cadence of 1 minute. 
See Section \ref{sec:101031} for details. 
(An animation of this figure is available.)} 
   \label{f:mr-cc-101031}
   \end{figure}

\begin{figure}
   \centering
  \includegraphics[width=\textwidth, angle=0]{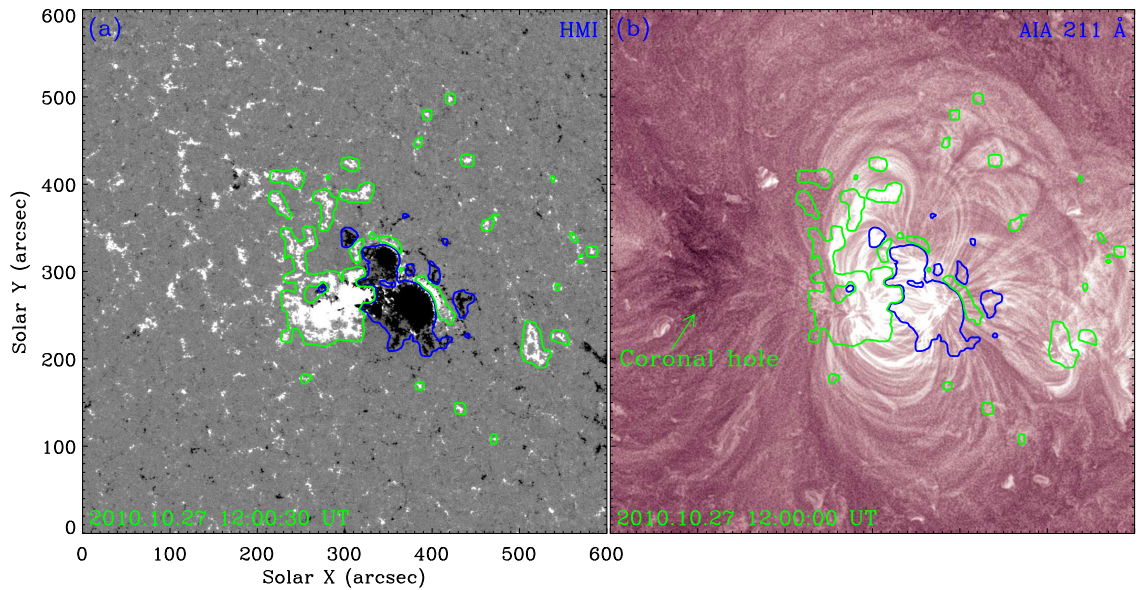}
   \caption{Similar to Fig.\,\ref{f:mf-140501}, but for the on-disk magnetic fields and context coronal structures on October 27 of the coronal rain event on October 31 \citep{2011AA...532A..96K}, 2010. 
See Section \ref{sec:101031} for details. } 
   \label{f:mf-101031}
   \end{figure}

\begin{figure}
   \centering
  \includegraphics[width=\textwidth, angle=0]{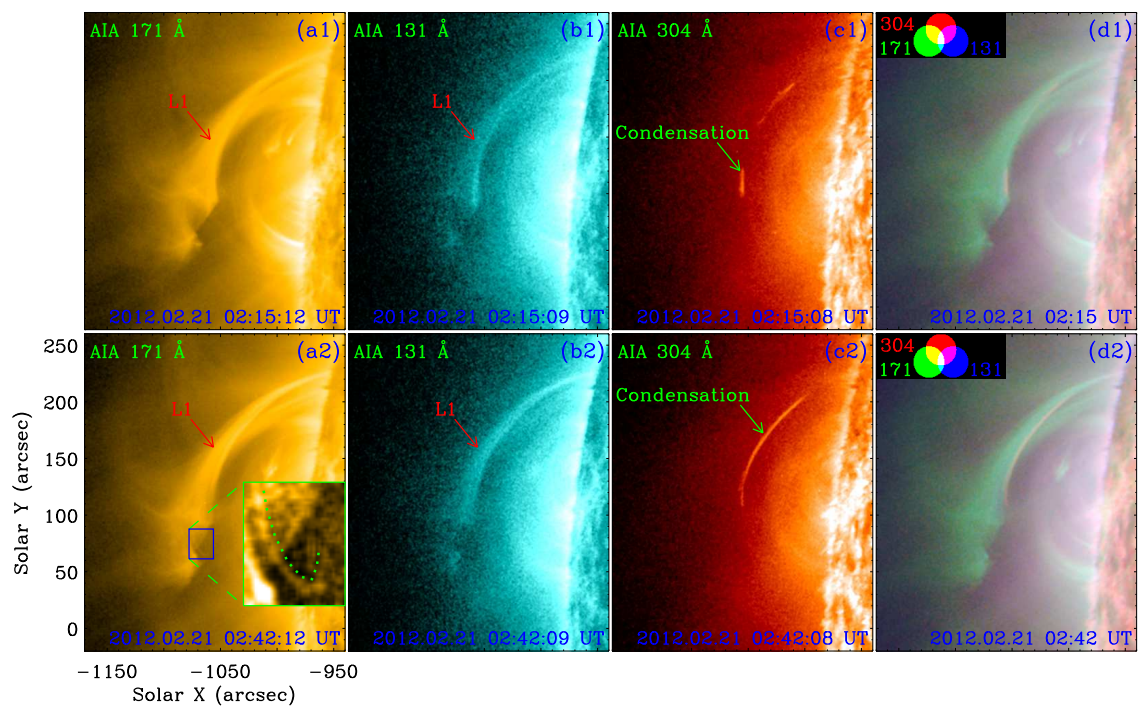}
   \caption{Similar to Fig. \ref{f:mr-cc-101126}, but for the coronal rain event on 2012 February 22 \citep{2015AA...577A.136V}.
In (a2), the AIA 171 \AA~image in the blue rectangle is enlarged in the green rectangle, and has been enhanced using the MGN technique.
The green dotted line in the green rectangle in (a2) outlines the newly reconnected closed loops. 
An animation of the unannotated AIA images is available online. 
It covers 5 hr starting at 00:00 UT on 2012 February 21, with a time cadence of 1 minute. 
See Section \ref{sec:120222} for details. 
(An animation of this figure is available.)} 
   \label{f:mr-cc-120222}
   \end{figure}

\begin{figure}
   \centering
  \includegraphics[width=\textwidth, angle=0]{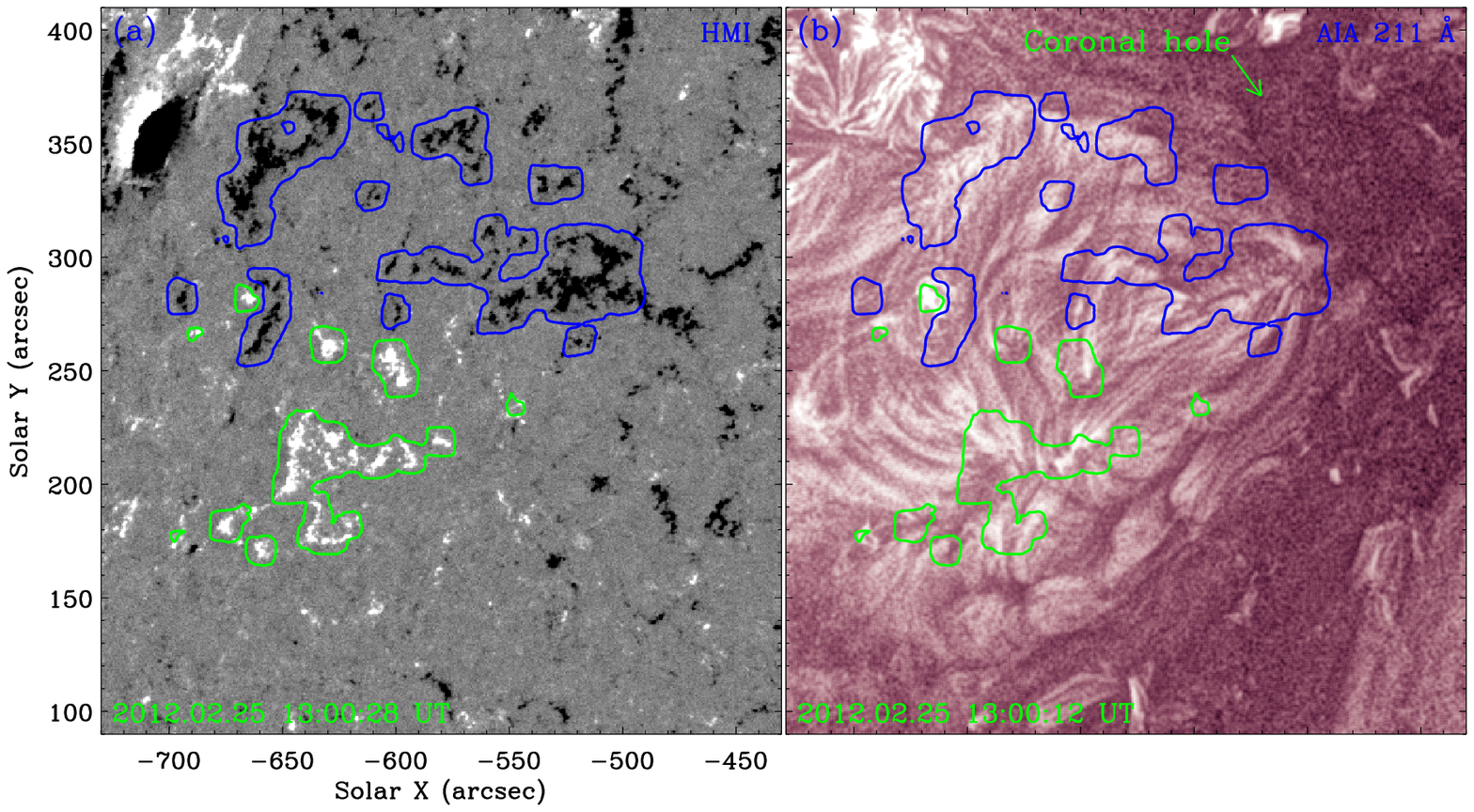}
   \caption{Similar to Fig. \ref{f:mf-140501}, but for the on-disk magnetic fields and context coronal structures on February 25 of the coronal rain event on February 22 \citep{2015AA...577A.136V}, 2012.  
See Section \ref{sec:120222} for details.} 
   \label{f:mf-120222}
   \end{figure}

\begin{figure}
   \centering
  \includegraphics[width=\textwidth, angle=0]{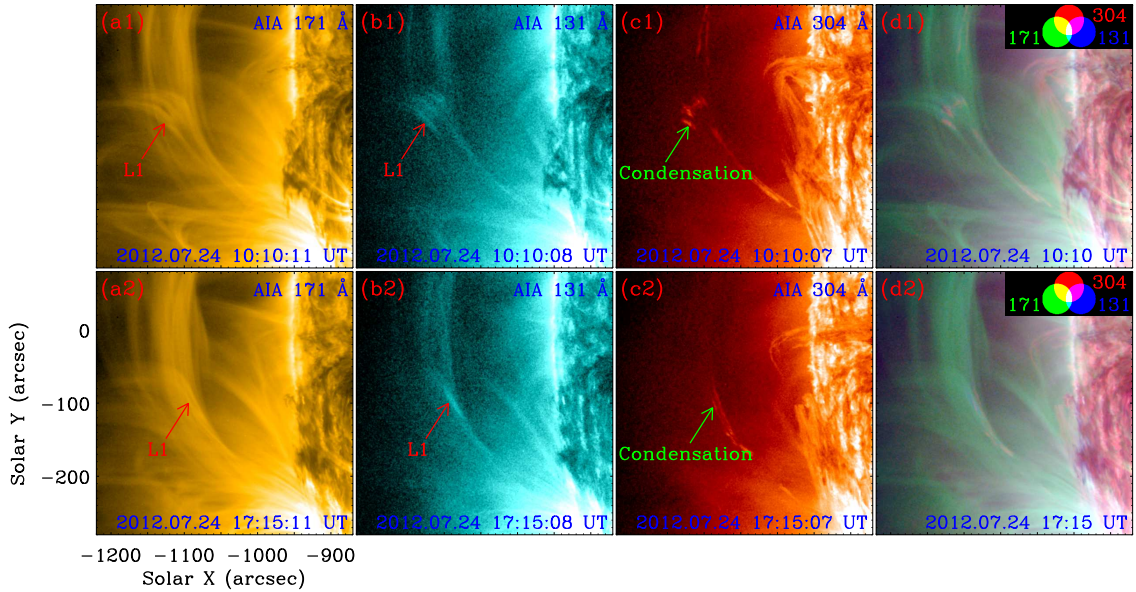}
   \caption{Similar to Fig. \ref{f:mr-cc-101126}, but for the coronal rain event on 2012 July 24 \citep{2018ApJ...853..176A}. 
An animation of the unannotated AIA images is available online. 
It contains data for two time periods that covers 3.5 and 2 hr starting at 08:00 and 16:00 UT on 2012 July 24, respectively, with a time cadence of 1 minute. 
See Section \ref{sec:120724} for details. 
(An animation of this figure is available.)} 
   \label{f:mr-cc-120724}
   \end{figure}

\begin{figure}
   \centering
  \includegraphics[width=\textwidth, angle=0]{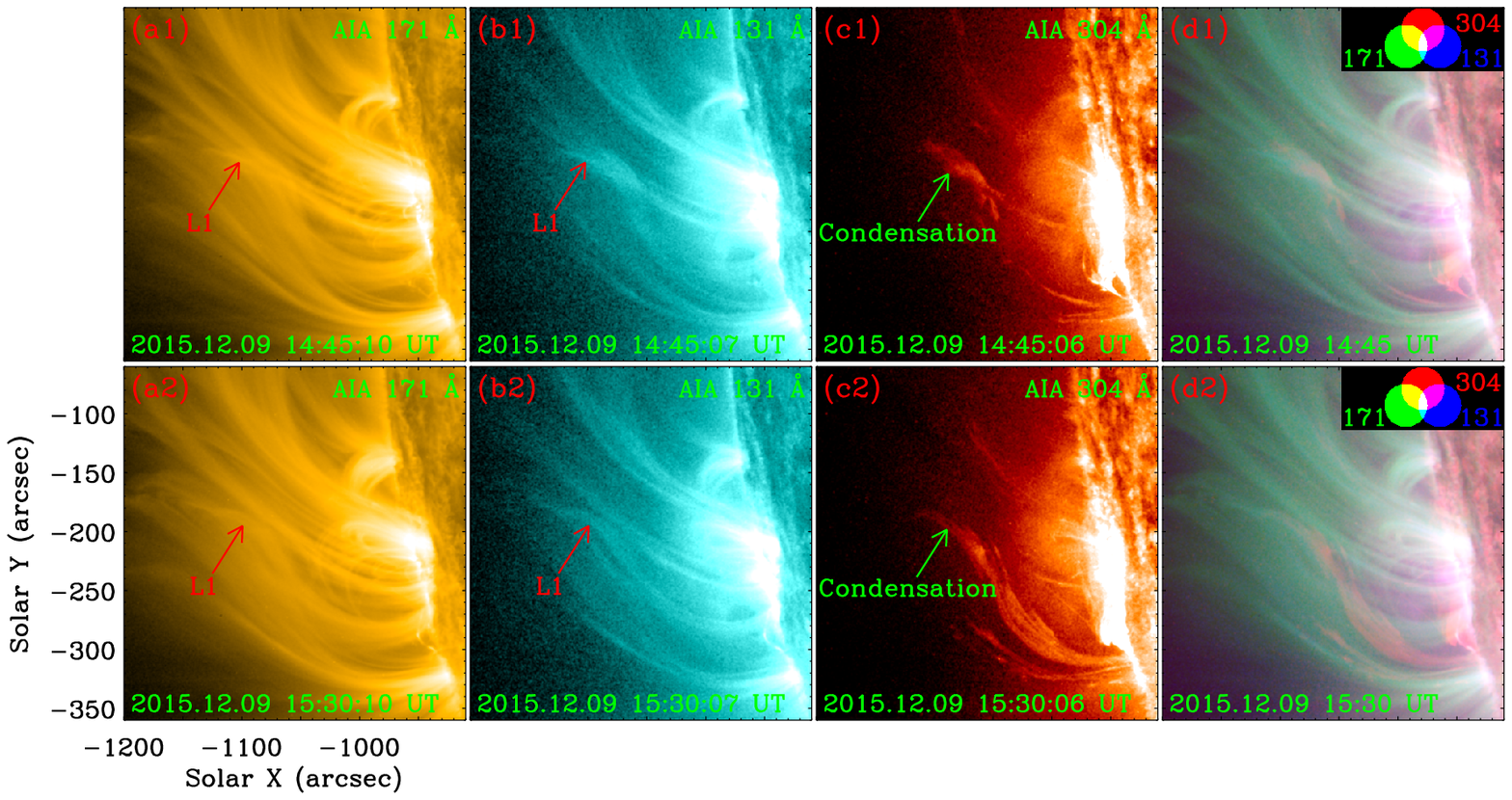}
   \caption{Similar to Fig. \ref{f:mr-cc-101126}, but for the coronal rain event on 2015 December 9 \citep{2017SoPh..292..132S, 2018ApJ...865...31S}. 
An animation of the unannotated AIA images is available online. 
It covers 5 hr starting at 13:00 UT on 2015 December 9, with a time cadence of 1 minute. 
See Section \ref{sec:151209} for details. 
(An animation of this figure is available.)} 
   \label{f:mr-cc-151209}
   \end{figure}

\begin{figure}
   \centering
  \includegraphics[width=\textwidth, angle=0]{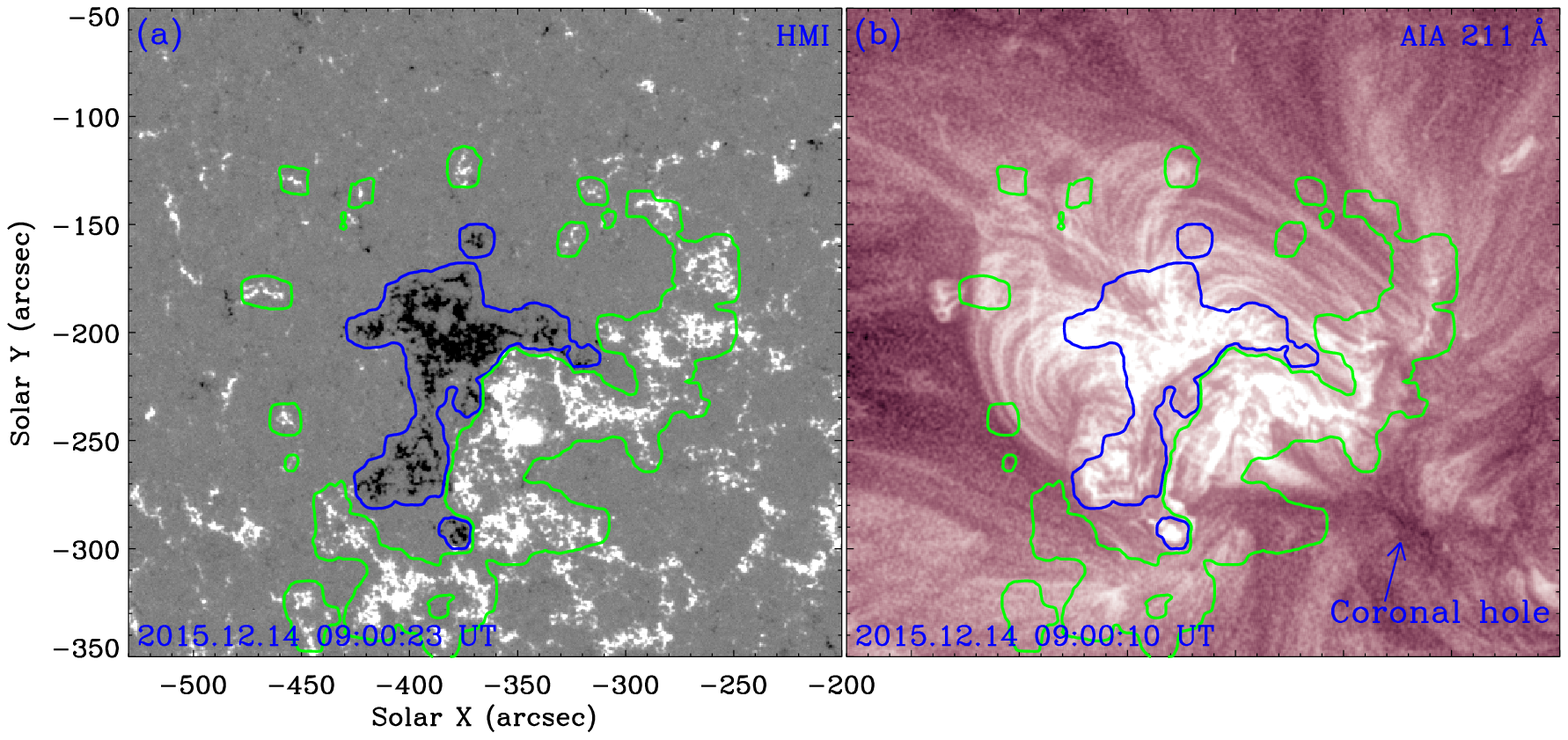}
   \caption{Similar to Fig. \ref{f:mf-140501}, but for the on-disk magnetic fields and context coronal structures on December 14 of the coronal rain event on December 9 \citep{2017SoPh..292..132S, 2018ApJ...865...31S}, 2015. 
See Section \ref{sec:151209} for details.} 
   \label{f:mf-151209}
   \end{figure}

Interchange reconnection at the null point of RNPT, facilitating the quiescent coronal rain along open outer spines, is identified.
Revisiting the coronal rain event in \citet{2015ApJ...807....7R}, the RNPT is detected in the source AR; see  Figs.\,\ref{f:mr-140501}-\ref{f:mf-140501}.
Examining the evolution of context coronal structures of the coronal rain event, interchange reconnection between the outer open spines and inner closed loops well under the fan surface, and formation of the newly reconnected outer open spines and inner closed loops, are clearly detected; see the online animated version of Fig.\,\ref{f:mr-140501}.
This supports observationally the FM for quiescent coronal rain along open outer spines in ARs via interchange reconnection at the null point \citep{2019ApJ...874L..33M}.
Moreover, interchange reconnection between the open structures from the AR and the closed loops in the quiet Sun region surrounding the AR is observed; see Sections\,\ref{sec:101126} and \ref{sec:120724}, besides that at the null point of RNPT in the ARs. 
This is consistent with the FM for quiescent coronal rain facilitated by interchange reconnection between open and closed structures in the quiet Sun regions \citep{2018ApJ...864L...4L, 2018ApJ...868L..33L, 2019ApJ...884...34L, 2020ApJ...905...26L, 2021ApJ...910..82L}.

Two types of quiescent coronal rain are observed simultaneously in one coronal rain event.
Revisiting the coronal rain events in \citet{2018ApJ...853..176A} and \citet{2017SoPh..292..132S, 2018ApJ...865...31S}, part of the coronal rain event occurring along open structures is found in the south of the transequatorial closed loops from which most of the coronal rain event originates; see Sections \ref{sec:120724} and \ref{sec:151209}.
It takes place repeatedly along the open structures. This repetition may make a contribution to the periodic coronal rain in the transequatorial closed loops.
Here, the global topological changes caused by interchange reconnection clearly play a role in this repetition, as the repetition observed in the coronal rain along open structures closely follows the formation of dips \citep{2020ApJ...905...26L, 2021ApJ...910..82L}.
Even though they happen respectively along the closed and open field lines, both types of quiescent coronal rain play an elemental role in the mass cycle between the hot, tenuous corona and cool, dense chromosphere. 
They are commonly observed and repeatedly occurring phenomena \citep{2018ApJ...853..176A, 2019ApJ...884...34L, 2020ApJ...905...26L, 2020AA...633A..11F}.
For further understanding their FMs, more efforts are needed in the future.

\begin{table}
\bc
\begin{minipage}[]{\textwidth}
\caption[]{General information of the coronal rain events from previous studies. \label{tab}}\end{minipage}
\setlength{\tabcolsep}{1pt}
\small
 \begin{tabular}{|c|c|c|c|c|c|c|c|c|c|}
  \hline
No. & References & Date & Source & Coordinate & Position & Observations & Original & SDO & Revised \\
 & & & region & & & & FM & &  FM \\
  \hline
1 & \citet{2021ApJ...910..82L}  & 2011.07.14 & Near & S40 W02 & On-disk & SDO; & FM-II-2 & Yes & - \\
 & & -07.15 & AR 11250 & & Off-limb & STEREO & & & \\  
\hline
2  & \citet{2021ApJ...907...41K}  & 2015.04.20 & Plage & N07 E90 & Off-limb & SDO & FM-II-1;   & Yes & - \\
\cline{3-5}
 & & 2015.04.21 & AR 12333 & N20 E90 &  & & FM-II-2 & &  \\
\cline{3-5}
 & & 2015.01.09 & AR 12261 & S11 E90 &  & & & & \\
\hline 
3 & \citet{2020ApJ...905...26L}  & 2013.10.18 & Quiet Sun & S45 E90 & Off-limb & SDO; & FM-II-2 & Yes & - \\
 & & -10.19 &   & & &  IRIS  & & & \\  
\hline
4  & \citet{2020SoPh..295...53I} & 2014.04.24 & AR 12042 & N22 W32 & On-disk & SDO; IRIS & FM-II-1 & Yes & -  \\
\hline
5 & \citet{2020AA...633A..11F} & 2017.08.28 & AR 12674 & N10 E90 & Off-limb & SDO; & FM-II-1 & Yes & FM-II-1\\
 & & -08.30 &   & & & SST   & & & \\  
\hline
6 & \citet{2019ApJ...884...34L} & 2012.01.16 & Quiet Sun & N50 W90 & Off-limb & SDO; & FM-II-2 & Yes & - \\
 & & -01.20 & & & & STEREO & & & \\  
\hline
7 & \citet{2019AA...630A.123K} & 2015.12.09 & AR 12468 & S13 E90 & Off-limb & SDO; IRIS & FM-II-1 & Yes & FM-II-1  \\
\hline
8 & \citet{2019ApJ...874L..33M} & 2015.04.16 & Sharp 5437 & N00 W90 & Off-limb & SDO & FM-II-1; & Yes & - \\
\cline{3-5}
 & & 2015.05.04 & AR 12333 & N22 W90 &   &  & FM-II-2 & &  \\  
\cline{3-5}
 & & 2016.02.27 & AR 12488 & N04 W90 &   & & & & \\
\hline
9 & \citet{2019Ap.....62...69V} & 2011.10.06 & Near & N40 E90 & Off-limb & SDO & FM-II-1 & Yes &  FM-II-1 \\
 & & -10.07 & AR 11312 & & & & & &  \\  
\hline
10 & \citet{2018ApJ...864L...4L, 2018ApJ...868L..33L} & 2012.01.19 & Quiet Sun & N50 W90 & Off-limb & SDO & FM-II-2 & Yes &  - \\
\hline
11 & \citet{2017SoPh..292..132S, 2018ApJ...865...31S} & 2015.12.09 & AR 12468 & S00 E90 & Off-limb & SDO; & FM-II-1 & Yes & FM-II-1;\\
 &  & & & & & IRIS; DST & & &   FM-II-2 \\
\hline  
12 & \citet{2018ApJ...853..176A} & 2012.07.23 & AR 11532 & S00 E90 & Off-limb & SDO & FM-II-1 & Yes & FM-II-1;\\
 & & -07.25 & & & & & & &  FM-II-2 \\  
\hline
13 & \citet{2017ApJ...842...15L} & 2015.03.11 & AR 12297 & S10 E20 & On-disk & SDO; IRIS & FM-I & Yes & - \\
\hline
14 & \citet{2017AA...601L...2V} & 2014.08.27 & AR 12141 & N20 W90 & Off-limb & SDO; IRIS; & FM-II-1 & Yes & FM-II-1 \\
 & & & & & & Hinode & & & \\  
\hline
15 & \citet{2017AA...598A..57V} & 2012.04.16 & AR 11461 & N17 W90 & Off-limb & Hinode & FM-II-1 & Yes & FM-II-1 \\
\hline  
16 & \citet{2016ApJ...833..184S, 2014ApJ...797...36S}  & 2012.07.02 & AR 11515 & S17 E14 & On-disk & SDO; SST & FM-II-1 & Yes & - \\
\cline{3-10}
 & & 2012.07.01 & AR 11515 & S20 E23 & On-disk & SDO; & FM-I & Yes & - \\
\cline{3-5}
 & & 2011.09.24 & AR 11302 & N10 E60 & &  SST & & & \\
\hline
17 & \citet{2016ApJ...827...39K} & 2014.08.25 & Near & S13 E90 & Off-limb & SDO; IRIS; & FM-II-1 & Yes & FM-II-1 \\
 & & & AR12151 & & & Hinode & & & \\  
\hline
18 & \citet{2015ApJ...807....7R} & 2014.05.01 & AR 12044 & S20 W90 & Off-limb & SDO; IRIS & - & Yes & FM-II-2 \\
\hline
19 & \citet{2015ApJ...806...81A} & 2010.06.26 & AR 11084 & S20 E72 & Off-limb & SST; SDO & FM-II-1 & Yes & FM-II-1 \\
\cline{3-7}
 & & 2013.11.29 & AR 11903 & S15 W90 & Off-limb & SDO; IRIS; &   &  &  \\
 & & & & & & Hinode & & & \\
\hline
20 & \citet{2015AA...577A.136V} & 2012.02.22 & AR 11420 & N10 E90 & Off-limb & SDO;  & FM-II-1 & Yes & FM-II-2  \\
 & & & & & & STEREO & & & \\
\hline
21 & \citet{2015ApJ...803...85L} & 2014.05.09 & AR 12051 & S17 W90 & Off-limb & SDO; IRIS & FM-II-1 & Yes & FM-II-1 \\
\hline
22 & \citet{2014SoPh..289.4117A} & 2011.09.29 & AR 11305 & N12 E14 & On-disk & SDO; NST; & FM-II-1 & Yes & - \\
 & & & & & & STEREO & & & \\
\hline
23 & \citet{2012ApJ...745L..21L} & 2010.11.26 & AR 11126 & S32 W90 & Off-limb & SDO & - & Yes & FM-II-2 \\
\hline
 \end{tabular}
\ec
\end{table}

\setcounter{table}{0}
\begin{table}
\bc
\begin{minipage}[]{\textwidth}
\caption[]{(Continued) }\end{minipage}
\setlength{\tabcolsep}{1pt}
\small
 \begin{tabular}{|c|c|c|c|c|c|c|c|c|c|}
  \hline
No. & Reference & Date & Source & Coordinate & Position & Observations & Original & SDO & Revised \\
 & & & region & & & & FM & &  FM \\
  \hline
24 & \citet{2012SoPh..280..457A} & 2008.06.10 & AR 10998 & S07 E63 & On-disk & SST & FM-II-1 & No & - \\
\cline{3-5}
 & & 2008.06.11 & AR 10998 & S09 E46 &  & & & & \\
\cline{3-9}
 & & 2010.06.27 & AR 11084 & S19 E65 & On-disk  & SST & FM-II-1 & Yes & \\
\cline{3-5}
 & & 2010.06.28 & AR 11084 & S19 E45 &  & & & & \\
\cline{3-5}
 & & 2010.07.06 & AR 11084 & S19 W60 & & & & & \\
\hline
25 & \citet{2012ApJ...745..152A} & 2009.05.10 & AR 11017 & N18 E90 & Off-limb & SST; & FM-II-1 & No & -  \\
 & & & & & &  Hinode & & & \\
\hline
26 & \citet{2011AA...532A..96K} & 2010.10.31 & AR 11117 & N20 W90 & Off-limb & SDO & FM-II-1 & Yes & FM-II-2 \\
\cline{3-10}
 & & 2010.11.05 & AR 11120 & N38 W09 & On-disk & SDO; & FM-II-1 & Yes & - \\
 & & & & & &  Hinode & & & \\
\hline 
27 & \citet{2010ApJ...716..154A}; & 2006.11.09 & AR 10921 & S06 W80 & Off-limb & Hinode & FM-II-1 & No & -   \\
 & \citet{2011ApJ...736..121A}  & & & & & & & & \\ 
\hline
28 & \citet{2009RAA.....9.1368Z} & 2007.05.10 & Quiet Sun & S02 E90 & Off-limb & STEREO; & FM-II-1 & No & - \\
 & & & & & &  Hinode & & & \\
\hline
29 & \citet{2007AA...475L..25O} & 2003.03.21 & AR 10314 & S13 W90 & Off-limb & TRACE; & FM-II-1 & No & - \\
 & & & & & & SOHO  & & & \\
\hline
30 & \citet{2004AA...415.1141D, 2005AA...443..319D}; & 2001.07.11 & AR 9538 & N17 E90 & Off-limb & SOHO; & FM-II-1 & No & -  \\
 & \citet{2005AA...436.1067M} & & & & & BBSO & & & \\
\hline
31 & \citet{2003ApJ...586.1417B} & 2001.06.15 & AR 9502 & S26 E48 & On-disk & SOHO & FM-I & No & -  \\
\hline
32 & \citet{2001SoPh..198..325S} & 2000.05.26 & AR 9004 & N11 W83 & Off-limb & TRACE & FM-II-1 & No & - \\
\cline{3-5}
 & & 1999.05.29 & AR 8531 & N17 E90 &  & & & & \\
\hline
\end{tabular}
\ec
\tablecomments{\textwidth}{In columns 8 and 10, the FM-I denotes the FM for flare-driven coronal rain, and the FM-II-1 and FM-II-2 represent the FM for quiescent coronal rain separately along the closed and open field lines. 
The dashed lines in column 8 show that no FM for coronal rain was clearly provided, and those in column 10 indicate that the coronal rain events are not revisited. 
See Section \ref{sec:res} for more details.}
\end{table}

\normalem
\begin{acknowledgements}
The authors thank the anonymous referee for helpful comments. We are indebted to the SDO team for providing the data.
This work is supported by the Strategic Priority Research Program of Chinese Academy of Sciences, grant No. XDB 41000000, the National Natural Science Foundations of China (12073042 and 11873059), the Key Research Program of Frontier Sciences (ZDBS-LY-SLH013) and the Key Programs (QYZDJ-SSW-SLH050) of Chinese Academy of Sciences, 
and Yunnan Academician Workstation of Wang Jingxiu (No. 202005AF150025).
H.Q.S. is supported by the National Natural Science Foundation of China (U2031109). 
We acknowledge the usage of JHelioviewer software \citep{2017A&A...606A..10M} and NASA's Astrophysics Data System.

\end{acknowledgements}


\end{document}